\documentclass[final,5p,times,twocolumn,authoryear]{elsarticle}

\usepackage{amssymb}
\usepackage{lipsum}
\usepackage{booktabs}
\usepackage{float}
\usepackage{amsmath}
\usepackage{xcolor}

\journal{Astronomy $\&$ Computing}

\begin{document}

\begin{frontmatter}



\title{Emulators for stellar profiles in binary population modeling}

\author[1,2,11]{Elizabeth\, Teng}
\author[2,11,3]{Ugur\, Demir}
\author[1,2]{Zoheyr\, Doctor}
\author[2,11,3]{Philipp\,M.\,Srivastava}
\author[3]{Shamal\, Lalvani}
\author[1,2,11]{Vicky\, Kalogera}
\author[2,11,3]{Aggelos\, Katsaggelos}
\author[4,5]{Jeff\, J.\,Andrews}
\author[6,7]{Simone\,S.\,Bavera}
\author[6,7]{Max\, M.\,Briel}
\author[2,11]{Seth\, Gossage}
\author[8,9]{Konstantinos\, Kovlakas}
\author[6,7]{Matthias\,U.\,Kruckow}
\author[1,2]{Kyle\, Akira\, Rocha}
\author[2]{Meng\, Sun}
\author[6,7]{Zepei\, Xing}
\author[10]{Emmanouil\,Zapartas}

\affiliation[1]{organization={Department of Physics and Astronomy, Northwestern University}, 
                addressline={2145 Sheridan Road}, 
                city={Evanston}, 
                state={IL},
                postcode={60208}, 
                country={USA}}
\affiliation[2]{organization={Center for Interdisciplinary Exploration and Research in Astrophysics, Northwestern University}, 
                addressline={1800 Sherman Avenue}, 
                city={Evanston}, 
                state={IL},
                postcode={60201}, 
                country={USA}}
\affiliation[11]{organization={NSF-Simons AI Institute for the Sky (SkAI)}, 
                addressline={172 E. Chestnut Street}, 
                city={Chicago}, 
                state={IL},
                postcode={60611}, 
                country={USA}}
\affiliation[3]{organization={Department of Electrical and Computer Engineering, Northwestern University}, 
                addressline={2145 Sheridan Road}, 
                city={Evanston}, 
                state={IL},
                postcode={60208}, 
                country={USA}}
\affiliation[4]{organization={Department of Physics, University of Florida}, 
                addressline={2001 Museum Road}, 
                city={Gainesville}, 
                state={FL},
                postcode={32611}, 
                country={USA}}
\affiliation[5]{organization={Institute for Fundamental Theory}, 
                addressline={2001 Museum Road}, 
                city={Gainesville}, 
                state={FL},
                postcode={32611}, 
                country={USA}}
\affiliation[6]{organization={Département d’Astronomie, Université de Genève}, 
                addressline={Chemin Pegasi 51}, 
                city={Versoix}, 
                postcode={CH-1290}, 
                country={Switzerland}}
\affiliation[7]{organization={Gravitational Wave Science Center (GWSC), Université de Genève}, 
                addressline={Chemin Pegasi 51}, 
                city={Versoix}, 
                postcode={CH-1290}, 
                country={Switzerland}}
\affiliation[8]{organization={Institute of Space Sciences (ICE), CSIC, Campus UAB}, 
                addressline={Carrer de Can Magrans s/n}, 
                city={Barcelona}, 
                postcode={E-08193}, 
                country={Spain}}
\affiliation[9]{organization={Institut d'Estudis Espacials de Catalunya (IEEC), Edifici RDIT, Campus UPC}, 
                city={Castelldefels (Barcelona)}, 
                postcode={E-08860}, 
                country={Spain}}
\affiliation[10]{organization={Institute of Astrophysics, Foundation for Research and Technology}, 
                addressline={N. Plastira 100}, 
                city={Heraklion}, 
                postcode={70013}, 
                country={Greece}}

\begin{abstract}
Knowledge about the internal physical structure of stars is crucial to understanding their evolution. The novel binary population synthesis code \texttt{POSYDON} includes a module for interpolating the stellar and binary properties of any system at the end of binary \texttt{MESA} evolution based on a pre-computed set of models. In this work, we present a new emulation method for predicting stellar profiles, i.e., the internal stellar structure along the radial axis, using machine learning techniques. We use principal component analysis for dimensionality reduction and fully-connected feed-forward neural networks for making predictions. We find accuracy to be comparable to that of nearest neighbor approximation, with a strong advantage in terms of memory and storage efficiency. 
By providing a versatile framework for modeling stellar internal structure, the emulation method presented here will enable faster simulations of higher physical fidelity, offering a foundation for a wide range of large-scale population studies of stellar and binary evolution.

\end{abstract}


\end{frontmatter}

\section{Introduction} \label{sec:intro}

    The study of binary evolution stands as an essential and complex area in astrophysics, offering a unique probe into the underlying physical processes governing the lives of most stars and their impact on their environments. Binary star systems serve as invaluable laboratories for testing theoretical models and advancing our understanding of stellar evolution and the production of compact objects and their transformation into electromagnetic and gravitational-wave sources. Furthermore, a large fraction of high-mass stars are born in binaries, making binarity a key factor in massive star evolution \citep[e.g.,][]{duchene, chini}.  A standard way to study the diversity of binary stars and their evolutionary paths is with population synthesis codes, which enable simulation of populations of stellar binaries\footnote{Although population synthesis methods have been used for electromagnetic sources for almost 40 years, the recent discoveries of compact binary gravitational-wave sources have led to renewed interest.}. This has been especially useful in recent years with LIGO-Virgo-KAGRA opening the door to a significant sample of gravitational-wave discoveries of compact object mergers \citep{gwtc1,gwtc2,gwtc3}. In comparing observed distributions of compact object merger properties like masses, spins, and redshifts, to simulated evolution outcomes of synthetic binary populations, we can better constrain uncertain stages like common envelope evolution and supernovae. 

    There is a broad slate of existing binary population synthesis codes, most of which rely on analytical fits of single-star evolution tracks based on \cite{hurley}, which fit to stellar evolution models computed by \citet{pols}. These include \texttt{SeBa} \citep{seba}, \texttt{binary\_c} \citep{izzard04,izzard06,izzard09}, \texttt{StarTrack} \citep{startrack}, \texttt{COSMIC} \citep{cosmic}, \texttt{COMPAS} \citep{compas},  and \texttt{MOBSE} \citep{mobse}. This approach is very computationally efficient as no actual binary evolution models are used and it is therefore widely used for population-level studies. However, it relies on a very old, outdated set of single-star, non-rotating models and require a large number of additional recipes to account for all the effects of binary evolution. These approximations can lead to conclusions that are inconsistent with population studies based on evolutionary models that self-consistently simulate evolution of the star's structure and binary interactions \citep[e.g.,][]{monica, klencki,patton}. More recent codes allow for additional flexibility in the adopted stellar physics by providing different choices of single-star evolutionary tracks as alternatives to those used in \citet{hurley}. 
    This has been done by using look-up tables and adding basic binary prescriptions, for example in \texttt{COMBINE} \citep{combine} and \texttt{SEVN} \citep{sevn}. Another approach is to self-consistently apply effects of mass transfer to single-star evolution, but only fully compute the evolution of one star at a time for computational constraints, as \texttt{BPASS} does \citep{bpass}. Most recently \citet{posydonv1} introduced \texttt{POSYDON}, a binary population synthesis code that represents the current state-of-the-art in fully self-consistently solving stellar structure equations for both stars concurrently with orbital evolution using the 1D stellar structure code \texttt{MESA} \citep{mesacode}. 

    The detailed stellar evolution simulations done with \texttt{MESA} are computationally expensive, with each binary evolutionary track sometimes taking over 100 CPU hours to run \citep{posydonv1}. Using low-cost machine learning models trained on smaller (tens of thousands of binaries) pre-computed \texttt{MESA} grids is one of the ways to efficiently generate large populations of binaries. These machine learning models emulate outcomes of \texttt{MESA} evolution \citep{posydonv1}. The first version of \texttt{POSYDON} includes a newly developed interpolation module for mass-transferring binaries at all evolutionary stages, which provides final values for scalar stellar and binary properties including mass, effective temperature, radius, and orbital period as a function of initial binary parameters of stellar mass and orbital period. At the end of each evolutionary stage, \texttt{POSYDON} also saves information about the full internal stellar structure of the final model generated as part of the \texttt{MESA} simulation. According to convention in stellar astrophysics, we use the term ``stellar profiles" to refer to the dependence of any physical quantity that varies along the interior of the star. 
    
    Stellar profiles contain valuable information that is not available from integrated or surface quantities \textemdash information which can be used to constrain later evolutionary stages and to investigate anomalous behavior. For example, the density profile of a star can be used to inform its potential core-collapse stage. 3D simulations have shown that the density profile of a star plays a major role in explodability. Early accretion through a large, sharp drop in the density profile can cause the confining ram pressure to abruptly decrease and the shock to rapidly expand, resulting in an explosion \citep{vartanyan18,vartanyan19,vartanyan21}. The outcome of this explosion has proven to be very sensitive to the location of the density jump \citep{luca23}.
    
    The density and angular rotation profiles of a progenitor can be used to estimate remnant properties. Using the mass and angular momentum distributions of the star, which can be calculated from these profiles, \citet{simone2020} add infalling shells of material to the black hole and associated accretion disk in a process described by \citet{batta}. The mass and angular momentum added from this material can be used to estimate the mass and spin of the resulting black hole throughout and after collapse. The distributions of predicted black hole spins can then be compared to observational distributions \citep{simone2020}.
    
    The chemical composition profiles can be used to calculate the core-envelope boundary, which can then be used to determine whether the system has enough energy to eject a common envelope, a critical phase for the formation of interacting binaries with compact objects \citep{dewi,ivanova,matthias2016}. 
    
     Having accurate emulated profiles readily available will open the door to many lines of investigation in binary population synthesis studies, especially when it comes to custom user-provided prescriptions for uncertain evolutionary stages. This is especially useful for cases where a user wants to test out a prescription on the fly without being limited by default settings or having to generate an entirely new set of grids. 
        
    Due to the correlations between the profile points in the stellar profile dataset, profiles cannot be interpolated in the same way that the global quantities like stellar mass are in the existing interpolation module of \texttt{POSYDON}, which is described in Section \ref{sec:dens} of this paper. For that reason, we present here a new methodology to emulate stellar profiles, which uses machine learning techniques to both reduce the dimensionality and provide full predictions of stellar profiles for key physical quantities. This is a flexible approach which aims to provide a starting point for efficient emulation of stellar structure. While demonstrated here with density profiles from our \texttt{POSYDON} simulations, our framework can be applied to any physical property of interest.
    
    For the predictions, we leverage neural networks, a powerful and readily available tool commonly employed to learn complex patterns and relationships in large datasets \citep[e.g.,][]{aggelos}. Specifically, we utilize fully-connected neural networks due to their superior ability to approximate functions and because they evaluate very quickly after training, which is essential for large-scale applications of function approximation problems such as in binary population synthesis. Previous work has successfully used neural networks to emulate stellar evolution simulation outputs, including single-star evolutionary tracks \citep{kiril}, fundamental stellar properties of $\delta$ Scuti stars \citep{astero}, and observable stellar properties \citep{lyttle}. More broadly, the use of machine learning methods in general to emulate various aspects of stellar evolution has grown in recent years, including the use of normalizing flows to emulate single-star evolutionary grids \citep{flow}, augmenting such grids with Gaussian process regression \citep{li}, and mapping observable quantities of stars to fundamental stellar parameters using random forests \citep{bellinger}.

    In Section \ref{sec:data} we describe the data used to train our profile emulators, and in Section \ref{sec:methods} we explain the machine learning techniques used in making predictions. We present measurements of model accuracy and discuss our results in section \ref{sec:results}. 

\section{Data} \label{sec:data}

    In this work we develop profile emulators trained on the \texttt{POSYDON} v1 binary \texttt{MESA} grids\footnote{Available at zenodo.org/records/6655751} 
    which are described in \citet{posydonv1}. These are: (i) a grid of interacting binaries with two hydrogen-rich main-sequence stars (HMS-HMS), (ii) a grid with one compact object\textemdash neutron star or black hole\textemdash and one hydrogen-rich star at the onset of Roche-lobe overflow (CO-HMS), and (iii) a grid with one compact object and one He-rich star (CO-HeMS). 

    \begin{table*}[!t]
        \centering
        \begin{tabular}{llll}
        \toprule
        & HMS-HMS & CO-HMS & CO-HeMS \\
        \midrule
        Number of Binary Simulations (Training Set) & 48,145 & 11,130 & 30,611 \\
        Number of Binary Simulations (Testing Set) & 2977 & 1661 & 2656 \\
        Percentage of Training Binaries with Stable Mass Transfer & 30.1 & 62.1 & 21.0 \\
        Percentage of Training Binaries with Unstable Mass Transfer & 35.8 & 32.4 & 3.1 \\
        Percentage of Training Binaries with No Mass Transfer & 25.8 & 0.0 & 61.6 \\
        Initial Primary Mass Range & 6.23\ -\ 120 $M_{\odot}$ & 0.5\ -\ 120 $M_{\odot}$ & 0.5\ - \ 80 $M_{\odot}$ \\
        Initial Primary Mass Values & 52 & 40 & 40 \\
        Initial Binary Mass Ratio Range & 0.05 \ - \ 1  & -& - \\
        Initial Binary Mass Ratio Values & 20 & -& - \\
        Initial Compact Object Mass Range & - & 1\ - \ 35.88 $M_{\odot}$ & 1\ -\ 35.88 $M_{\odot}$ \\
        Initial Compact Object Mass Values & - & 21 & 21 \\
        Initial Orbital Period Range & 0.7\ - \ 6105 days & 1.26\ - \ 3162 days & 0.02 \ - \ 1117.2 days \\
        Initial Orbital Period Values & 56 & 30 & 47 \\
        \bottomrule
        \end{tabular}
        \caption{\texttt{POSYDON} version 1 binary star grid properties. The percentages shown of each grid's mass transfer classes do not add up to 100 because we do not perform interpolation on binaries that are either non-physical or experienced numerical convergence errors. Note: in practice, we use a maximum mass ratio of 0.99 to avoid numerical issues for stars that initiate mass transfer at exactly the same time}
        \label{tab:grid_properties}
    \end{table*}

   The properties of the three \texttt{POSYDON} binary grids are listed in Table \ref{tab:grid_properties}, and more explanation is provided in \citet{posydonv1}. Grid values for initial primary masses, compact object masses, and orbital periods are spaced logarithmically, whereas initial binary mass ratio values for the HMS-HMS grid are evenly distributed in linear space\footnote{This distribution is somewhat modified for the CO-HMS grid during post-processing because the ``initial" binary parameters are replaced with the binary parameters at the onset of Roche-lobe overflow, as described in \citet{posydonv1}.}. While the v1 grids capture evolution at solar metallicity, \texttt{POSYDON} version 2 will include grids for 8 discrete metallicities: $2Z_{\odot}$, $Z_{\odot}$, $0.45 Z_{\odot}$, $0.2 Z_{\odot}$, $10^{-1}Z_{\odot}$, $10^{-2}Z_{\odot}$, $10^{-3}Z_{\odot}$, $10^{-4}Z_{\odot}$ \citep{posydonv2}. 

\begin{figure}[ht!]
    \includegraphics[width=\columnwidth]{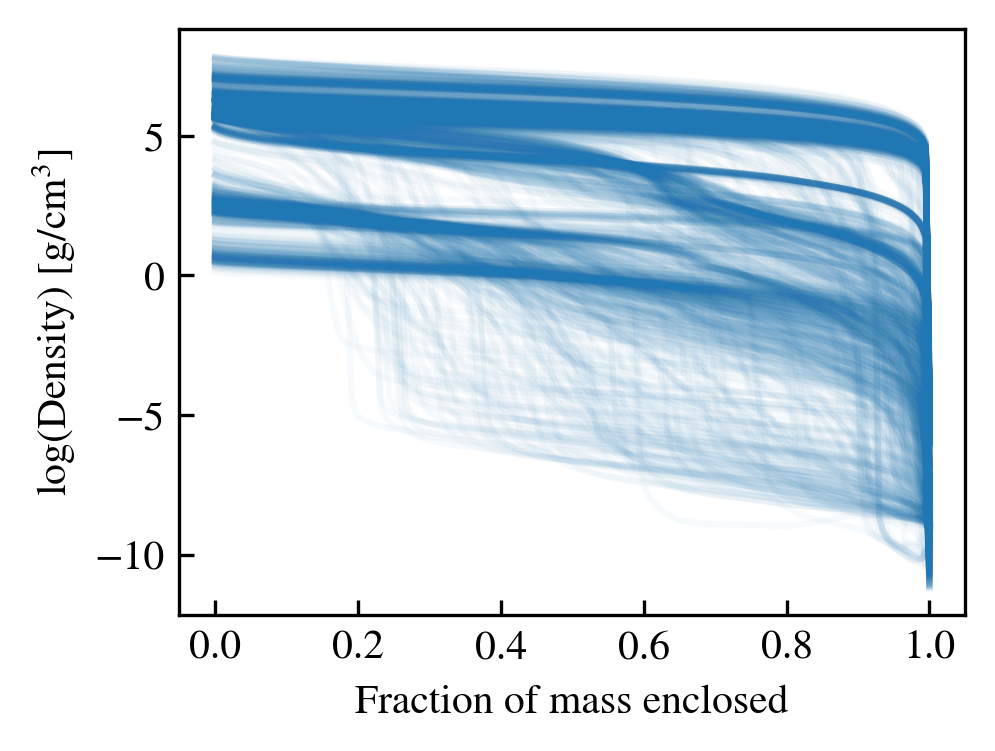}
    \caption{Log density profiles of all 1661 CO-HMS testing set binaries at the end of \texttt{MESA} evolution. The three main groups of central densities are hydrogen-burning stars at log(density)$\simeq$1, helium-burning stars at log(density)$\simeq$3, and carbon-burning stars at log(density)$\simeq$5. The highest density values are stars at carbon depletion, which is the latest stage at which we stop \texttt{MESA} simulations.}
    \label{fig:density}
    \end{figure}

    To evaluate the performance of interpolators trained on each of these \texttt{MESA} grids, we have three corresponding testing grids of binaries that are generated by random sampling of the same parameter spaces as the main binary grids. The end of \texttt{MESA} evolution in our models occurs when a star reaches the end of core carbon burning or the binary enters a common envelope phase (dynamically unstable mass transfer); more details about termination conditions can be found in Section 5.2 of \citet{posydonv1}. The outputs of these \texttt{MESA} grids include various properties of the binary at the end of \texttt{MESA} evolution, such as the binary orbital period, information about stellar structure, type of mass transfer experienced if any, and integrated or surface quantities of the stars, such as mass, radius, effective temperature, and luminosity. \texttt{POSYDON} also saves the stellar profiles at the end of \texttt{MESA} evolution, which show the structure of quantities like log density, log temperature, and hydrogen mass fraction at discrete sampling points throughout the star from the center to the surface as a function of either radius or mass. The placement of profile points is dictated by the adaptive mesh in the \texttt{MESA} model; points are more densely sampled in regions where higher resolution is necessary to characterize the complex behavior exhibited in the star \citep{mesa}. The specific quantities saved after a \texttt{MESA} run can be specified by the user generating their own grids; the default settings are explained in detail in \citet{posydonv1}. As an example of the profiles generated by \texttt{MESA}, in Figure \ref{fig:density} we show the final log density profiles of the binaries in the CO-HMS testing grid. 
    
    \subsection{Pre-processing}
    
        In this work, we express stellar profiles as a function of the mass enclosed. Final profiles provided by \texttt{MESA} sample points irregularly along the star's radial axis. We need the profile data to be uniformly sampled in order to work in our predictive models, so we use linear interpolation to resample each profile at 200 points evenly spaced along the enclosed mass axis. While in theory this results in loss of detail, in practice we find that the profiles do not have appreciable changes at the lengthscale of 1/200th of the mass enclosed. We experimented with doubling the resolution to instead resample each profile at 400 evenly spaced points. To compare the accuracy of these two resolutions, we linearly upsampled the 200-point profiles up to 400 points and calculated the relative error $\varepsilon_i$ at each profile point $i$ for every profile between the two profiles versions, one that was resampled to 400 directly from the MESA original, and one that was resampled to 200 from the original and then upsampled to 400:
        \begin{equation}
            \varepsilon_i =\left|\frac{\rho_{200}(m_{enc,i})-\rho_{400}(m_{enc,i})}{\rho_{400}(m_{enc,i})}\right|, 
        \end{equation}
        where $\rho_{200}(m_{enc,i})$ and $\rho_{400}(m_{enc,i})$ are respectively the upsampled and resampled density values of the $i$-th profile point, both with 400 points. We found that for the HMS-HMS, CO-HMS, and CO-HeMS grids respectively, 98.0\%, 97.8\%, and 98.8\% of all profile points have a relative error below 1\% between the resampled and upsampled versions. Our choice of resolution must balance accuracy with compactness, and we determined that increasing the resolution beyond 200 samples does not improve the profiles enough to be justifiable.

\section{Methods} \label{sec:methods}

    \texttt{POSYDON} v1 grids sample the parameter space across both stellar masses and the binary orbital period. Our goal is to predict profiles (physical quantity vs. mass enclosed) at the end of \texttt{MESA} evolution for any binary in this parameter space.\footnote{Given predicted density profiles, we can also express these profiles post hoc as a function of radius.} 

    While the fully-connected neural networks we use for emulation are less interpretable than, for example, tree-based methods, the high accuracy, scalability and computational efficiency of evaluation provide a good fit for our profile emulation task. Neural networks have previously been used by \citet{kiril} to efficiently emulate single-star evolutionary tracks calculated with \texttt{MESA}. This track emulation application, like the profile emulation presented here, is very useful for population synthesis studies. Neural network emulation has also been applied to many other astrophysical simulations, including by \citet{Ribas}, who use neural networks to emulate synthetic protoplanetary spectral energy distributions at any point in their model's parameter space. In this application, neural networks make it possible to efficiently sample posterior distributions for inference on protoplanetary disk masses, which would not be possible with the full radiative transport modeling. \citet{cosmonet} likewise use neural networks to emulate cosmic microwave background power spectra to ultimately conduct parameter estimation on cosmological parameters. 
    
    Gaussian process regression is sometimes used in emulation of astrophysical model outputs, for example observable stellar properties from single-star \texttt{MESA} simulations \citep{li}, stellar spectra \citep{czekala}, and cosmological simulations \citep{habib}. However, Gaussian processes are not well suited for high-dimensional problems or large datasets due to the large amount of memory required in computing the covariance matrix \citep{gpml}. 

    Datasets with high-dimensional outputs can necessitate more complex neural network architecture as well as more data. Either including a more complex model architecture or using more training data will also make the model more expensive to optimize \citep{stat}. Our profiles are 200-dimensional, and we did find that reducing the number of dimensions resulted in better predictions. This provides a more compact representation, so that the neural networks can train and make profile predictions in a lower-dimensional space. This is possible because we do not need 200 dimensions to express most of the variance in our dataset, as will be shown in Figure \ref{sec:dimred}. 

    To reduce our profiles' dimensionality we use principal component analysis (PCA), which is a common technique in astrophysics. PCA provides a representation space in which data points are most widely distributed by choosing axes along which the dataset displays the most variance \citep{PCA}. The first principal component is along the direction with maximal variance, the second principal component is orthogonal to the first along the direction that maximizes the residual variance in the dataset, and so on. 

    \subsection{Emulating density profiles}\label{sec:dens}

        The first release of \texttt{POSYDON} includes a module for interpolating binary evolution outcomes of scalar properties at the end of MESA evolution, including the star's final central density, for any binary in the parameter space covered by the training grids. This scalar initial-final interpolator, which is described in Section 7 of \citet{posydonv1}, uses either nearest-neighbor or linear interpolation depending on the user's specification. We need to ensure that the profile predictions are consistent with other interpolated values. We also want to avoid having the network redundantly learn these values. For these reasons, we apply minimum-maximum normalization to each profile, simplifying the prediction task to the profile shape:

        \begin{equation} \label{eq:minmax}
            \mathrm{normalized \: profile} = \frac{\mathrm{profile}-\mathrm{profile \:min}}{\mathrm{profile \:max}-\mathrm{profile \:min}}. 
        \end{equation}

        In this context, the `profile' refers to the array of values describing a given stellar property (e.g., density) as a function of the fraction of mass enclosed, and the `profile min' and `profile max' denote the minimum and maximum values of this array, respectively.
        
        Clearly, the values in each normalized profile range from zero to one. We make profile predictions in this normalized space, which are later scaled up using values from \texttt{POSYDON}'s scalar initial-final interpolation.
        
        \begin{figure}[ht!]
        \includegraphics[width=0.9\columnwidth]{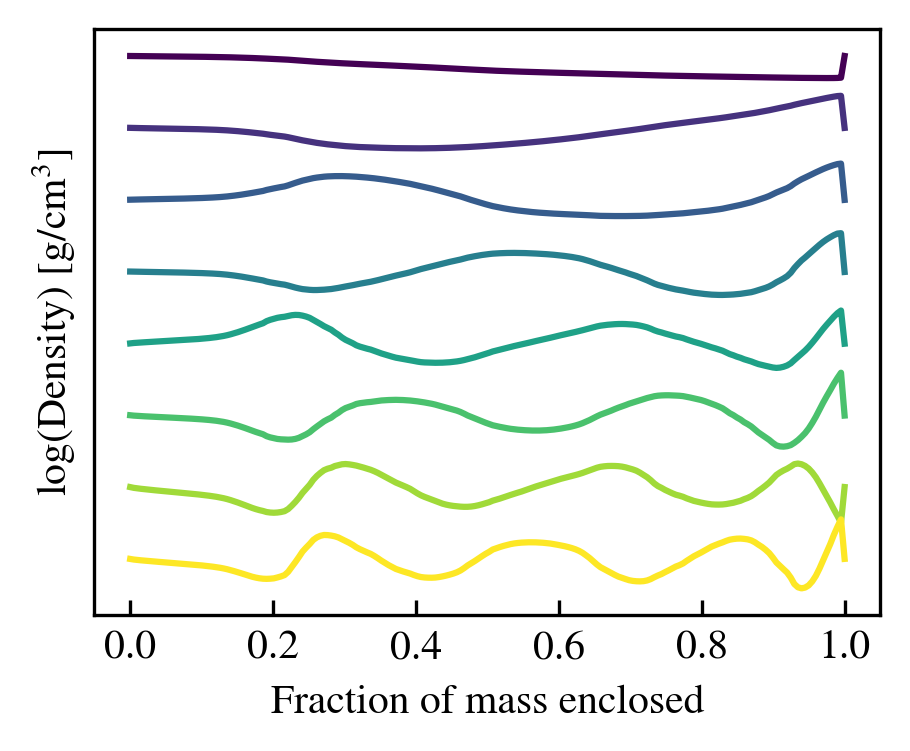}
        \centering
        \caption{Principal components of CO-HMS grid density profiles plus offsets for visual distinction. The first eight components are shown plotted in order of explained variance from top to bottom.}
        \label{fig:pcacomp}
        \end{figure}

        \subsubsection{Dimensionality reduction}\label{sec:dimred}
            We first reduce the dimensionality of our normalized profiles to simplify the task assigned to the neural network. The smoothly-varying nature of stellar quantities like density and temperature make PCA a suitable method. 
            In this process, each density profile, originally represented by 200 discrete data points sampled along the `fraction of mass enclosed' axis (shown in Figure \ref{fig:density}), is transformed into a lower-dimensional latent space. This latent space is spanned by a set of orthogonal 200-dimensional vectors, called principal components, which are derived from the training set and capture directions of maximum variance in the data. 
            The first eight principal components of density profiles from the CO-HMS grid, which account for the most variance in those profiles, are shown in Figure \ref{fig:pcacomp}. Each principal component contains 200 points along the same `fraction of mass enclosed' axis as the original profiles. The y-axis scaling of the principal components is arbitrary, as they represent orthogonal directional vectors in the data space; for clarity, we have vertically offset them in the figure for visual distinction. 
            Each training profile is projected onto the set of principal components to obtain the projection weights, which are used to create a reconstruction of the original profile as a weighted sum of the principal components. 
            
            One of the critical model design choices when using PCA is the number of principal components to keep. We need to balance compactness of the representation with fully capturing the dataset. A common metric for choosing the number of components is the cumulative variance explained by the first $n$ components. This is calculated from the eigenvalues of the covariance matrix of the data. The covariance matrix $C$ has dimensions $N \times N$, where $N$ is the total number of features in the original dataset, in this case 200. Each eigenvalue $\sigma_i$ represents the variance captured by the $i$-th principal component; the set of eigenvalues $\{\sigma_i\}$ is obtained by performing an eigenvalue-eigenvector decomposition of $C$, which is factored into its eigenvectors (the directions of maximum variance) and eigenvalues (the amount of variance explained by those directions). The cumulative variance explained by the first $n$ components is then given by:  
            \begin{equation}
                \mathrm{Cumulative \ Variance \ Explained} = \frac{\sum_{i=1}^n \sigma_i}{\sum_{i=1}^N \sigma_i}, 
            \end{equation}
            where $\sigma_i$ is the $i$-th eigenvalue and $N$ is the total number of original features \citep{stat}. 
            
            \begin{figure}[ht!]
            \includegraphics[width=\columnwidth]{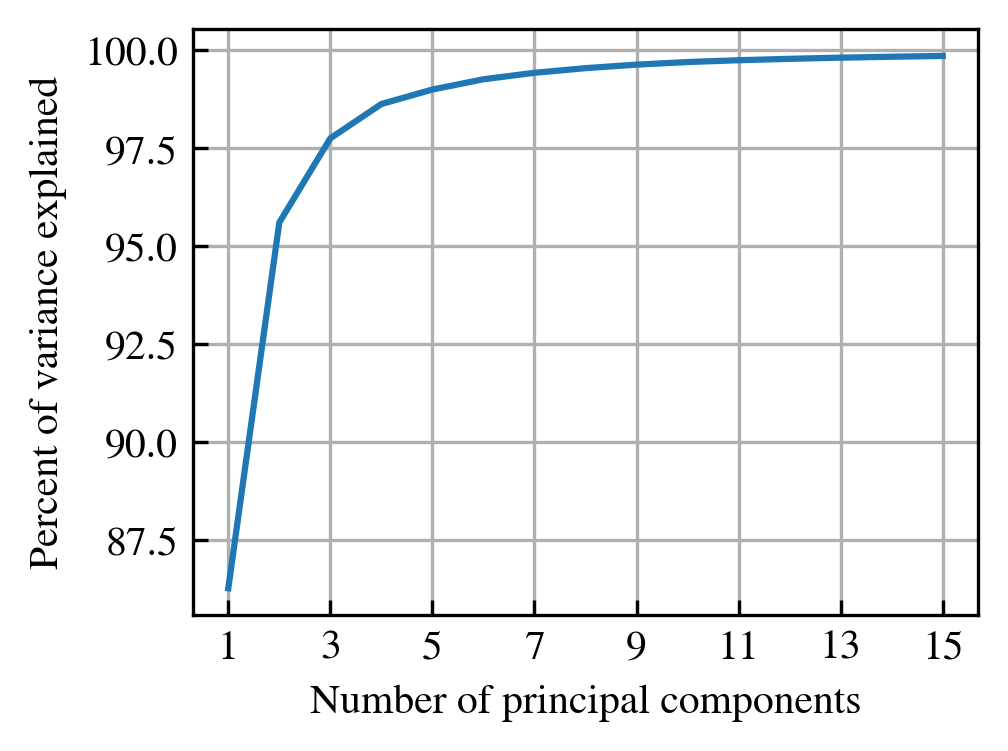}
            \caption{Percentage of dataset variance explained by principal components versus number of principal components included.}
            \label{fig:varexp}
            \end{figure}
            
            \begin{figure}[ht!]
            \includegraphics[width=\columnwidth]{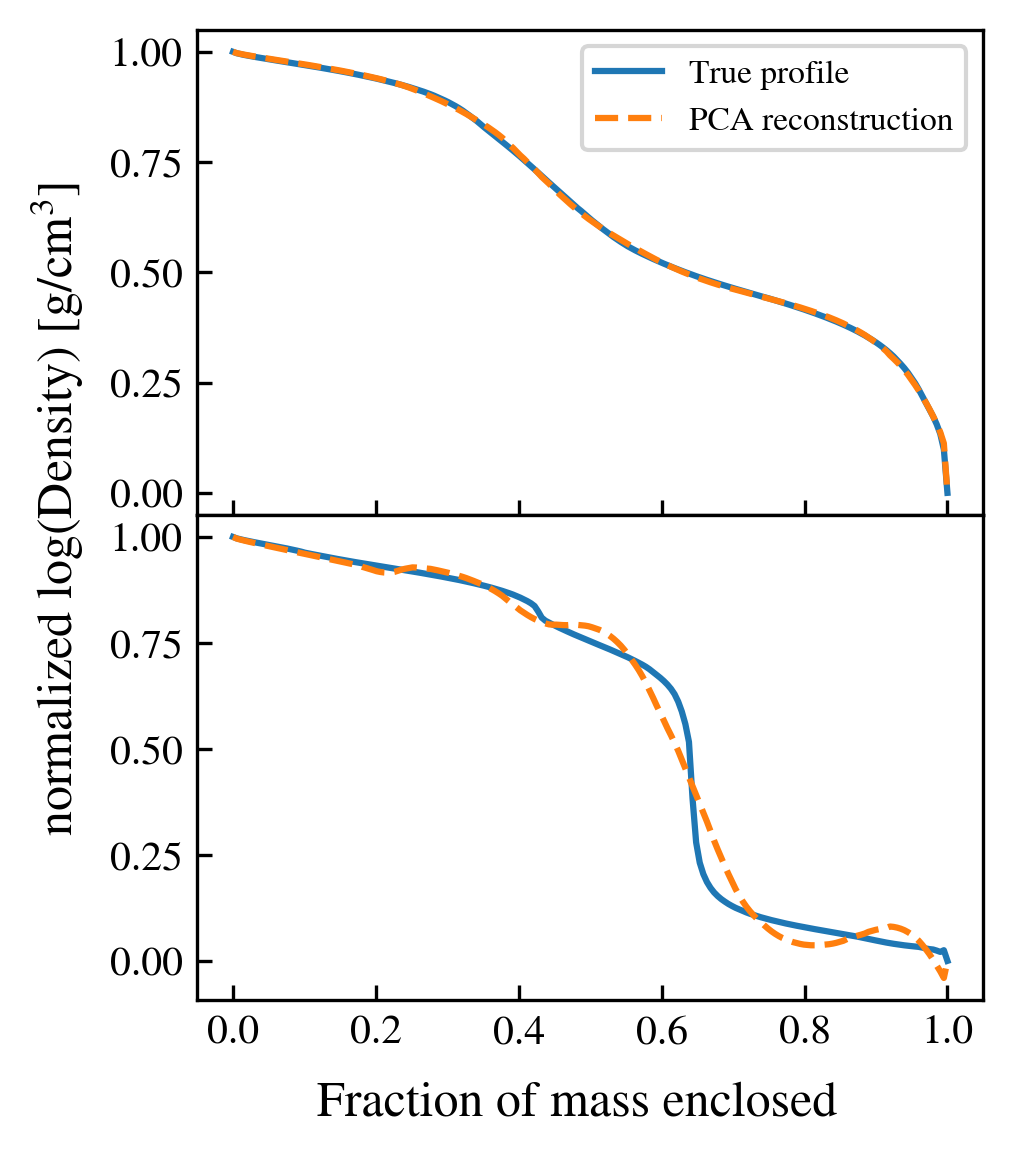}
            \caption{Two examples of PCA reconstructions (orange) of true density profiles (blue) shown for binaries in the HMS-HMS grid. The bottom panel shows a binary whose profile is in an area of the dataset's representation space not well covered by the principal component basis, and so the reconstructions in that area are less accurate than the example shown in the top panel.
            }
            \label{fig:compare_pca}
            \end{figure}
    
            We choose to include 8 components because at that point, the principal components explain ~99.5\% of the variance in the training dataset, and the variance added by additional components starts to level off, as shown in Figure \ref{fig:varexp}. It is worth noting that some profile shapes are much rarer than others in the training set used to generate principal components, so they are much more poorly matched by their PCA reconstructions, as shown in Figure \ref{fig:compare_pca}. We find that the most common shapes have near-constant density with a steep dropoff at the end, as seen in Figure \ref{fig:density}.  
        
            \begin{figure*}[ht!]
            \centering
            \includegraphics[width=1.5\columnwidth]{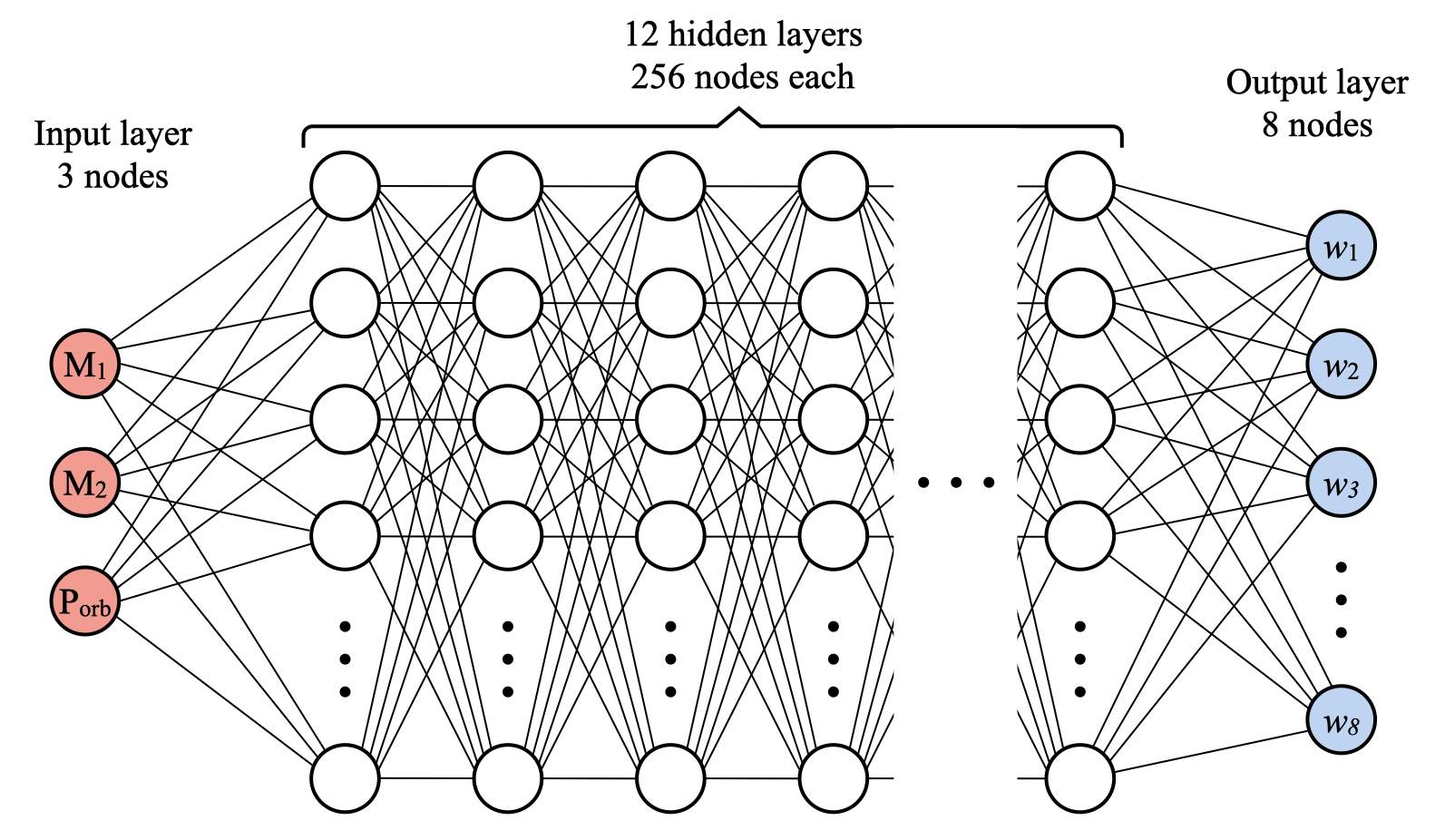}
            \caption{Neural network architecture for generating principal component weights of emulated density profiles. }
            \label{fig:nn}
            \end{figure*}
    
        \subsubsection{Network Architecture}
            Our emulator is based on fully-connected feed-forward neural networks, with the input layer consisting of three input nodes (for each of the dimensions of the \texttt{MESA} grids, which are the two stellar masses and orbital period). The networks have 8 output nodes for the profiles’ principal component weights. The depth and width of the network were chosen to balance model complexity and prediction accuracy with computational efficiency. 
            To do this, we conducted a grid search over various network sizes. The evaluation criteria for optimal performance were based on the percentage of median profile errors (explained in detail in Section \ref{sec:results}) below 10\% and below 100\%. Among the configurations that performed best under these metrics, we further analyzed the full error distributions and selected the architecture that visually demonstrated the most desirable overall error distribution. The configuration of 12 hidden layers with 256 nodes per layer, shown in Figure \ref{fig:nn}, was determined to be the most effective architecture for our study, providing the best overall performance across the combined testing data from all mass transfer classes in all three binary grids. We experimented with including normalization layers (such as batch normalization) to stabilize learning and potentially improve convergence, but we found that this led to worse performance, potentially having interfered with the model's ability to capture sufficient complexity. As a result, the normalization layers were excluded from the final architecture.
    
            We find that separating our data according to mass transfer class improves performance, so for each grid we produce one machine learning model for each mass transfer class. This is consistent with the existing initial-final interpolators from \texttt{POSYDON} v1 \citep{posydonv1}. For training each of these profile emulation models, we used an 80/20 split to produce a training set and validation set. This step ensures the model's ability to generalize outside the training data by identifying potential underfitting or overfitting. Each network was initialized randomly using the Glorot uniform kernel initializer and then given up to 3000 epochs to train to ensure the model had enough time to converge. However, to further prevent overtraining we used an ``early stopping" feature to stop training if the validation loss does not improve for 100 epochs. This number, called the ``patience”, was determined through experimentation to give the best results in optimization. 
    
            The models were optimized based on a mean squared error (MSE) loss function placing equal emphasis on each component's weight. 
            It may seem more intuitive to base the loss function on the accuracy of profile predictions rather than the weights, because the accuracy of the weights is incidental to the ultimate goal of profile emulation. However, in practice we found that using a loss function based on the reconstructed profiles rather than the component weights ultimately made a minimal difference to the prediction accuracy. Hence, we chose the MSE loss of the weights for its faster evaluation time. We chose the Adam optimization algorithm, which is a stochastic gradient descent method that dynamically adjusts learning rates for each parameter based on the gradient \citep{adam}. This makes it suitable for a wide range of tasks, including those with sparse gradients or complex loss surfaces. We found that Adam converged to a significantly lower loss compared to other gradient descent algorithms such as SGD \citep{sgd}, RMSprop \citep{rmsprop}, AdamW \citep{adamw}, and Adagrad \citep{adagrad}. We also found that Adam provided faster convergence and more stable training for our dataset and model architecture.
    
            We use the ReLU (Rectified Linear Unit) activation function for all of the hidden layers, which allows the networks to learn non-linear behavior in the training set. In our experiments, ReLU outperformed other activation functions in model accuracy, including tanh (hyperbolic tangent) and sigmoid, for which the models had trouble converging during the optimization process. We found that tanh and sigmoid constrained layer outputs too tightly, which reduced the models' ability to capture the complexity afforded by additional layers. By contrast, ReLU avoids these issues due to its simplicity and unbounded output for positive inputs, enabling the network to represent more complex patterns in the data. The output layer uses linear activation to ensure that final predictions for principal component weights are unconstrained. The model was implemented using TensorFlow \citep{tensorflow} using the Keras API \citep{chollet2015keras}.

            \begin{figure*}[ht!]
            \includegraphics[width=1.8\columnwidth]{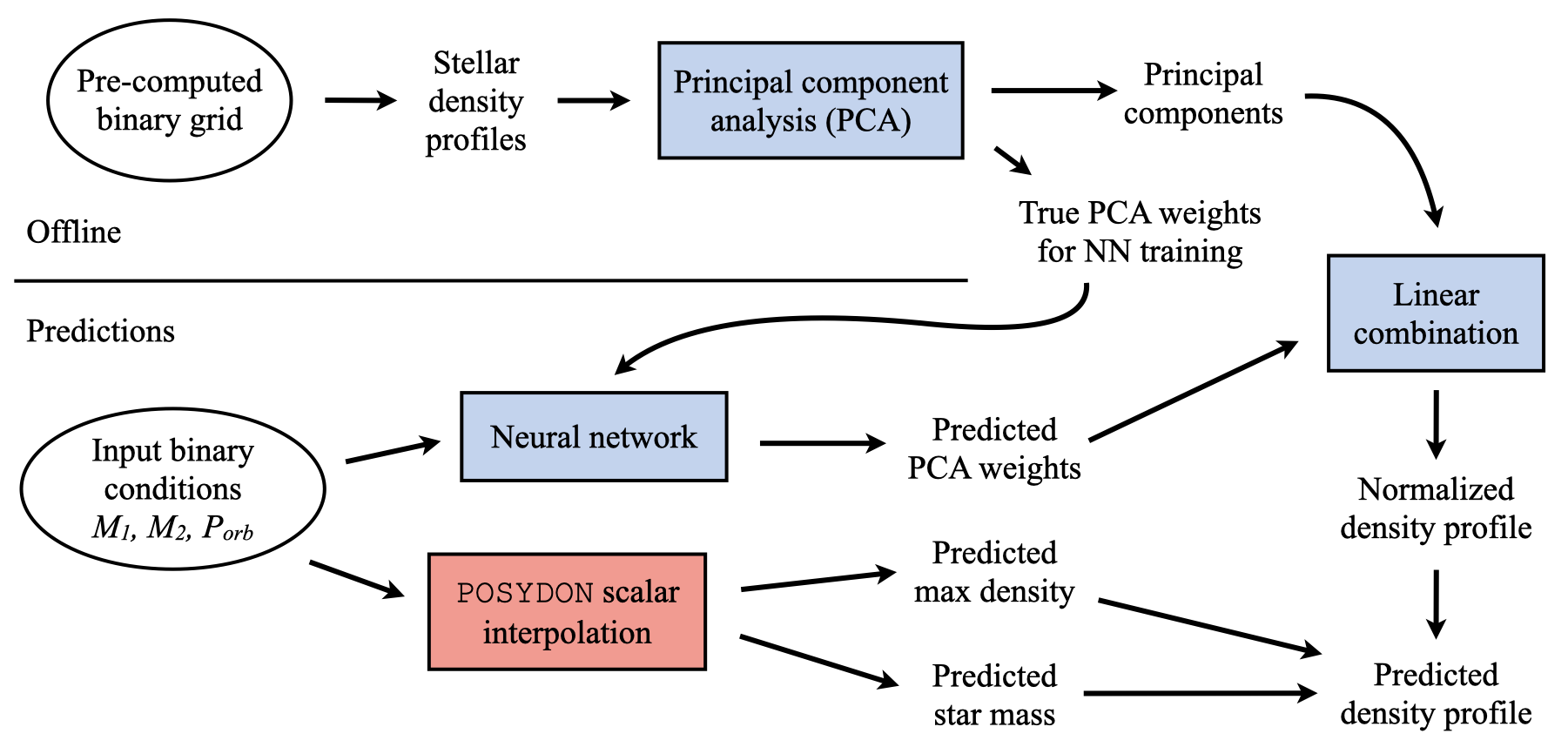}
            \centering
            \caption{Flowchart showing procedure for emulating density profiles. 
            The input binary conditions $M1, M2, P_{orb}$ represent the mass of the primary star, the mass of the secondary star, and the orbital period, respectively.} Red boxes are model components from \texttt{POSYDON}'s existing scalar interpolation module, blue boxes are the model components presented in this work, and white boxes are data products. \texttt{POSYDON}'s scalar interpolation uses the initial parameters and associated MESA grids to predict final scalar properties at the end of an evolutionary stage. For density profiles, the final mass and central log density of each star is predicted by this interpolator; the emulator presented here then produces a density profile which is re-scaled by these predicted values. 
            \label{fig:density_flowchart}
            \end{figure*}
        
            The overall procedure for obtaining predictions (illustrated in the flowchart in Figure \ref{fig:density_flowchart}) is as follows: we perform PCA on the training profiles to obtain a set of eight principal components and the eight principal component weights corresponding to each training profile. We then train neural networks with the training binaries' initial conditions (star 1 mass, star 2 mass, orbital period) as input and eight principal component weights as outputs. With this network in place, we predict the weights for the profiles of a star in any new binary, which combined with the principal components construct predicted normalized profiles. We then renormalize these profiles using values for central log density and final star mass predicted with \texttt{POSYDON}’s scalar interpolation module, which predicts these quantities from the initial conditions and the associated MESA grid. With this configuration and the design choices detailed above, we achieve sufficient performance for profile emulation. Future work experimenting with different choices or architectures may further improve accuracy.

        \subsubsection{Post-processing}
            \begin{figure}[ht!]
            \includegraphics[width=\columnwidth]{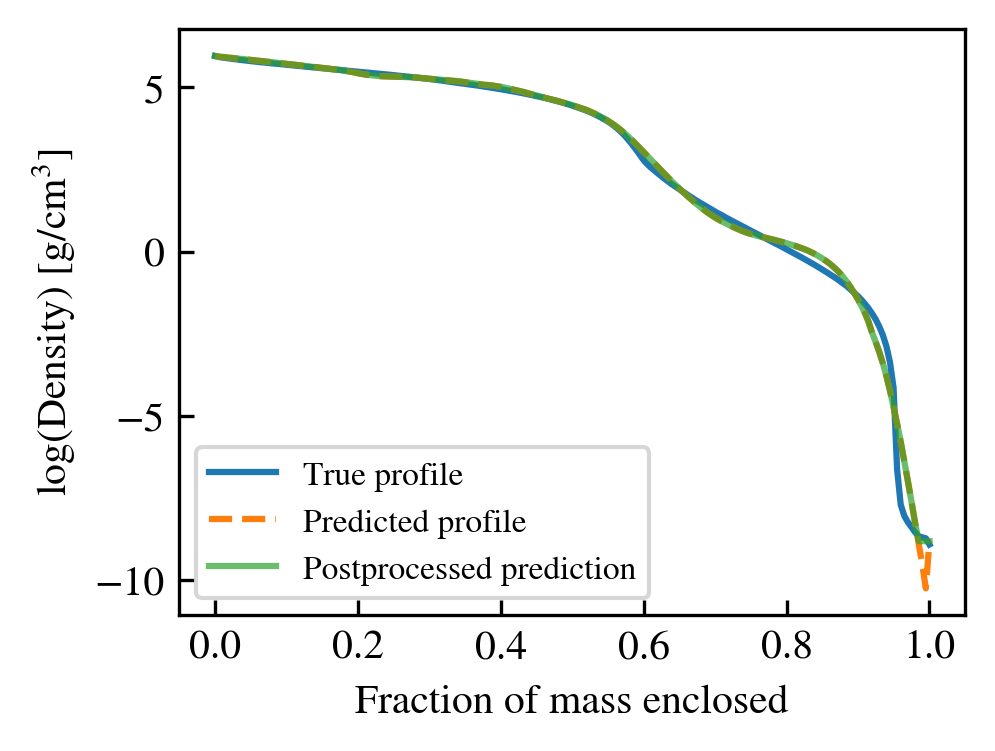}
            \caption{A predicted density profile in orange against the true profile in blue. The post-processed profile with monotonically decreasing density is over-plotted in green. This particular profile prediction from an HMS-HMS binary is at the 68th percentile for median profile error within its grid; the distribution is shown in Figure \ref{fig:HMSHMS_dens_CDF}.}
            \label{fig:mono}
            \end{figure}
    
            Some of the predicted profiles display slight sinusoidal behavior as a result of the sinusoidal shapes of the principal components; we post-process the predicted profiles to enforce monotonic decreasing behavior so that the stellar density profile decreases from center to surface with no density inversions for all profiles. We accomplish this by starting at the surface and setting the value of any point whose value is lower than the previous point to that of the previous point. Since the profiles are renormalized according to interpolated central density, we also cut off any values above that value. An example of a predicted density profile before and after enforcing monotonic decreasing is shown in Figure \ref{fig:mono}. Another option for implementing this would be to start at the center and force rising profile points downward; as it turns out both options produce near-identical error distributions against both the unprocessed predictions and the true profiles. 

\section{Results} \label{sec:results}

    In this section we share quantitative information on predictive model evaluation. To isolate the errors produced by the prediction methods presented in this work from those arising from the scalar interpolation methods published previously in \citet{posydonv1}, we use the true final central density values from the pre-computed test grid rather than interpolated values. This ensures that the errors evaluated in this study are solely attributable to the machine learning methods developed here, without influence from the scalar interpolation that would be used in practical profile emulation.

    We also compare the error distributions of our profile predictions to the profile of the nearest neighbor binary in the initial parameter space. Like with the predictions, we rescale these nearest neighbor profiles to the true central density values to provide an even comparison. 
    Many astrophysical applications of neural network emulators, such as \citet{Ribas} and \citet{cosmonet}, do not provide such evaluations against nearest neighbor approximation because they focus on calculating a posterior probability density and not on providing a surrogate for the original model. However, as presented in \citet{kiril}, nearest neighbor approximation is well-suited for population synthesis applications, so our emulator's predictions must compete with the accuracy of nearest neighbor profiles.

\begin{figure}[ht!]
    \includegraphics[width=\columnwidth]{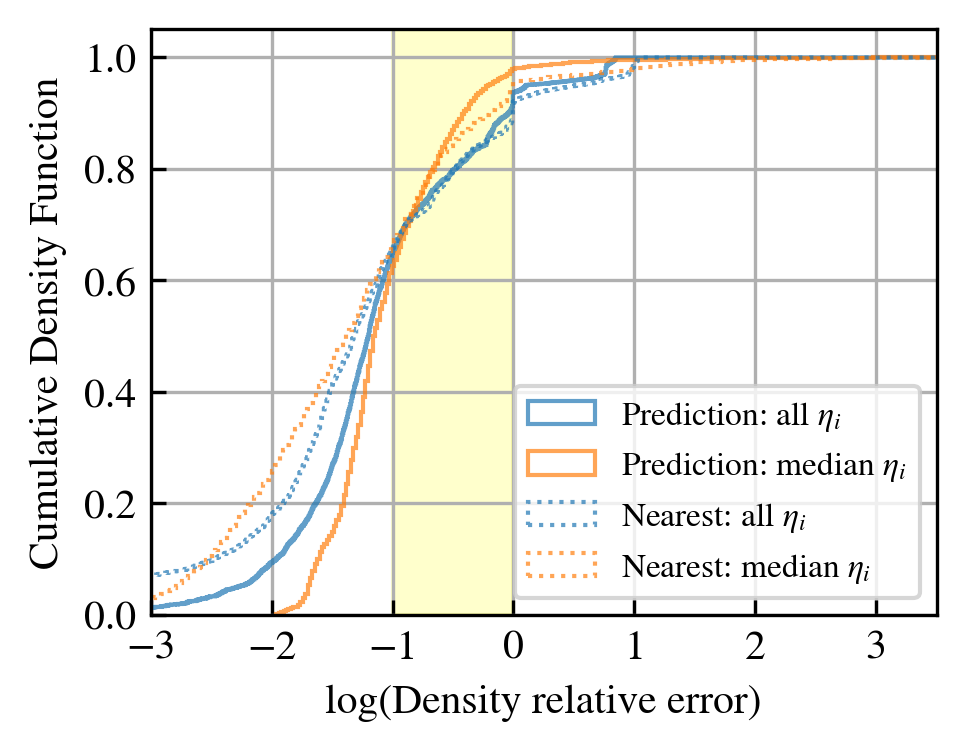}
    \caption{Cumulative density functions showing relative errors in HMS-HMS density profiles. We compare our predicted profiles in solid lines to rescaled nearest neighbor profiles in dotted lines. The distributions of all profile errors $\eta_i$ in the grid are in blue, whereas orange shows the distributions of the per-profile median $\eta_i$. The yellow highlighted area signifies the region of interest between 10\% and 100\% error. }
    \label{fig:HMSHMS_dens_CDF}
    \end{figure}

\begin{figure}[ht!]
    \includegraphics[width=\columnwidth]{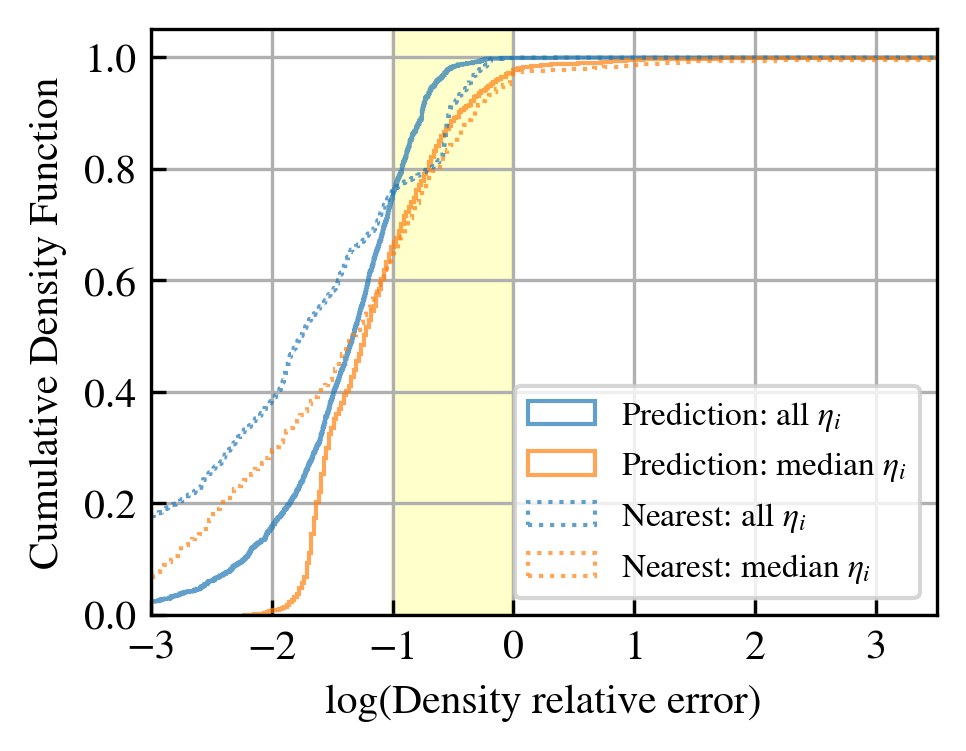}
    \caption{Cumulative density functions showing relative errors in CO-HMS density profiles. We compare our predicted profiles in solid lines to rescaled nearest neighbor profiles in dotted lines. The distributions of all profile errors $\eta_i$ in the grid are in blue, whereas orange shows the distributions of the per-profile median $\eta_i$. The yellow highlighted area signifies the region of interest between 10\% and 100\% error. }
    \label{fig:COHMS_dens_CDF}
    \end{figure}

\begin{figure}[ht!]
    \includegraphics[width=\columnwidth]{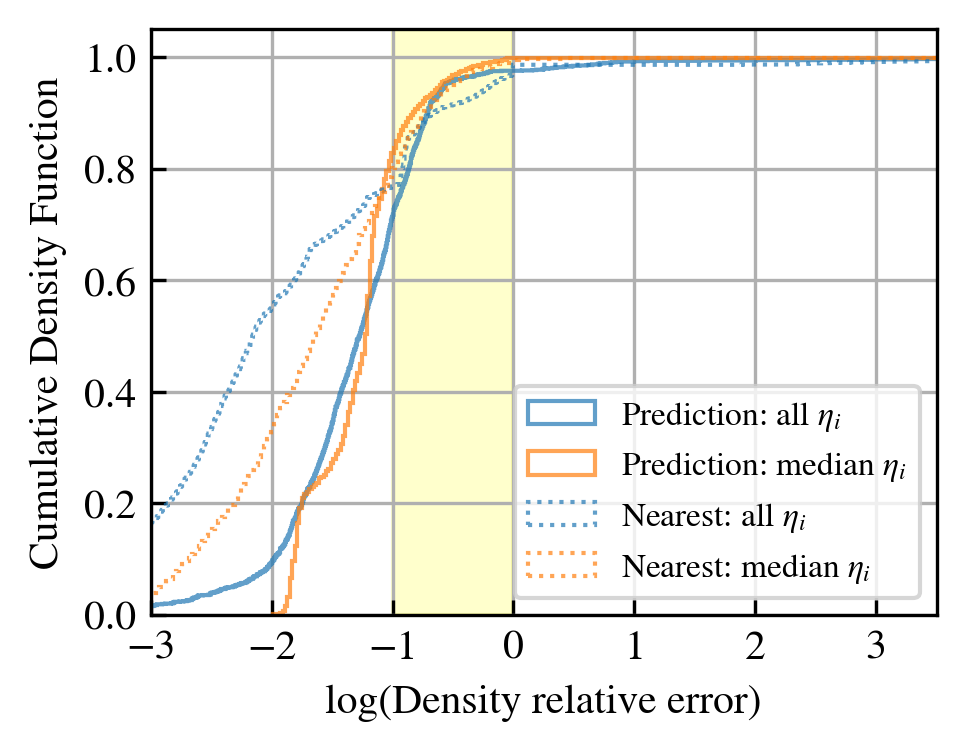}
    \caption{Cumulative density functions showing relative errors in CO-HeMS density profiles. We compare our predicted profiles in solid lines to rescaled nearest neighbor profiles in dotted lines. The distributions of all profile errors $\eta_i$ in the grid are in blue, whereas orange shows the distributions of the per-profile median $\eta_i$.The yellow highlighted area signifies the region of interest between 10\% and 100\% error. }
    \label{fig:COHeMS_dens_CDF}
    \end{figure}

    We examine the numerical error in the predicted density profiles. We measure the relative error $\eta_i$ of each profile point as:
    \begin{equation}
        \eta_i =\left| \frac{\rho_{pred}(m_{enc,i})-\rho_{true}(m_{enc,i})}{\rho_{true}(m_{enc,i})}\right|. 
    \end{equation}
    where $\rho_{pred}(m_{enc,i})$ and $\rho_{true}(m_{enc,i})$ are the predicted and true density values of the $i$-th profile point, respectively.  
    To measure the accuracy of the entire testing set, we consider the distribution of all $\eta_i$ for all profiles within the set. We also consider the median $\eta_i$ for each profile to be able to evaluate the accuracy of each profile. Both of these distributions are shown for each of the grids in Figures \ref{fig:HMSHMS_dens_CDF}, \ref{fig:COHMS_dens_CDF}, and \ref{fig:COHeMS_dens_CDF}, respectively. We use cumulative distribution functions (CDFs) to represent the error distributions. At each error value, the CDF is given by the fraction of errors in the distribution that fall below that value; essentially a histogram plotted cumulatively from lowest to highest error. When reading these plots to determine which method is more accurate, it is generally useful to look for which curve is farther to the left at a given percentile, signifying a lower error value at that percentile, or which curve is higher for a given error value, signifying that more errors in the distribution fall below that error value.

    When it comes to comparing the machine learning model's errors against nearest neighbor errors, we are most concerned with maximizing performance between 10\% and 100\% relative error. This is because improvements in performance below 10\%, for example from 1\% to 0.1\%, are negligible given the errors introduced by the code. When both the nearest neighbor profile and our predicted profile have errors below 10\%, the marginal difference in accuracy between the two methods is not significant for our purposes. Changes in errors above 100\%, for example between 200\% and 2000\%, are not as significant because 100\% is already a very significant error. Likewise in this region, when errors from both methods exceed 100\%, neither is usable, rendering any comparative advantage in accuracy irrelevant in such cases. 
    So we focus on the 10\% to 100\% range, where the comparison is consequential for our purposes. This region is highlighted in yellow in Figures \ref{fig:HMSHMS_dens_CDF}, \ref{fig:COHMS_dens_CDF}, and \ref{fig:COHeMS_dens_CDF}. When evaluating the CDF plots through this lens, the method with the better errors in this area of interest will have a distribution curve that is higher than the other curve in the area between -1 and 0 on the x-axis.
    
    Figure \ref{fig:HMSHMS_dens_CDF} shows that in the HMS-HMS grid, our best predictions are comparable to rescaled nearest neighbor profiles in the highlighted 10-100\% range. This grid also has the least favorable performance of the three compared to nearest neighbor, potentially due to the higher level of complexity in binary evolution compared to CO-HMS and CO-HeMS.  
        
    In Figure \ref{fig:COHMS_dens_CDF}, we show that our predictions in the CO-HMS grid somewhat outperform rescaled nearest neighbor in the highlighted 10-100\% relative error range. 
    
    We find that the CO-HeMS grid profiles are the least variable and complex out of the three grids, so there is the least room for improvement in the highlighted 10-100\% relative error range, as shown in Figure \ref{fig:COHeMS_dens_CDF}. Our machine learning predictions for this grid have a marginal improvement at most in this range. 
    

\section{Conclusions} \label{sec:conc}

    In this work, we present an emulation suite for the stellar profiles of binary evolution models produced with \texttt{MESA} and made public by the \texttt{POSYDON} collaboration. We use machine learning to predict the internal stellar structure for the evolution of interacting binaries. We find prediction accuracy to be comparable to (and in some cases better than) that of rescaled nearest neighbor approximation. However, machine learning emulators, like the one developed here, also provide the major advantage of memory and storage efficiency than nearest neighbor, since it only contains the neural network weights and principal components of the density profiles, rather than the full set of all the profiles in the grid. The three \texttt{POSYDON} v1 downsampled grid files are several gigabytes each, whereas the three profile interpolators are each under 100 megabytes. Additionally, the grid file size will increase as the number of binaries in the grid increases whereas the profile interpolator will remain more or less the same size regardless of grid size. This is especially relevant as the \texttt{POSYDON} project looks to expand the scope of the binary model grids to span multiple additional dimensions including metallicity in the forthcoming \texttt{POSYDON} version 2, and potentially eccentricity in the future. 
 
    
    Being able to emulate final stellar profiles at the end of \texttt{MESA} evolutionary stages in \texttt{POSYDON} has a number of potential science applications. Moreover, this study demonstrates the feasibility of a dimensionality reduction and neural network framework for stellar profile emulation in general. More information about the internal stellar profiles can provide insight into the behavior of binaries with anomalous properties. Being able to derive various physical quantities from emulated profiles instead of directly interpolating those quantities will expand the scope of binary population studies, particularly in cases where a target property was not saved in the original simulation. Emulated stellar profiles can also be used as an ingredient in custom physical prescriptions for defining the next steps of evolution, particularly in more uncertain phenomena such as common envelope evolution and supernovae. Rotation profiles in particular can also be leveraged to derive black hole spins of a population, which can be compared to observationally-derived distributions. By making our code publicly available, we aim to enable future studies to extend the approach presented here, paving the way for efficient and adaptable stellar profile emulators in diverse astrophysical contexts.

    The PCA-neural network method applied to density profiles here combines principal component analysis (PCA) for dimensionality reduction with neural networks to make predictions in the lower-dimensional latent space defined by the principal components, enabling efficient modeling of high-dimensional data. This approach is also well suited for other quantities with smooth profiles such as temperature.\footnote{Angular rotation profiles, despite appearing relatively smooth, do not monotonically decrease or have any consistency in normalized shape like density and temperature and therefore are not well represented by PCA. Without normalizing the profile, the PCA focuses on shapes of stars with very high internal rotation and neglects the profiles of more slowly rotating stars. Such profiles may require more elaborate neural network architecture without dimensionality reduction.} To improve emulation accuracy with this method, we could include more dimensions; our results with $n=8$ principal components are fairly successful without implementing any sophisticated model architecture. It would be possible to compensate for more principal components by optimizing the model further. 

    As discussed in Section \ref{sec:dens}, the performance of our method is affected by the fact that some of the density profile shapes are not well represented in the binary grids. This is because the training grids are constructed as rectilinear distributions that evenly sample the parameter space of binary initial conditions (as discussed in Section \ref{sec:data}). \citet{psycris} explores how active learning can be used to generate grids whose samples are optimally distributed for the user's advantage. They focus on optimizing grid sampling for combined mass transfer classification and scalar interpolation accuracy. The same method could also be used with a new loss function to optimize sampling for profile emulation; this would make the principal component basis for density profiles (along with other smooth-profile quantities) better at capturing rarer shapes, making prediction accuracy more evenly distributed across the grids. 

    Without having to change the original training grids, potential tactics to compensate for errors induced by the dataset's bias toward certain profile shapes could include adding in a classification step based on profile morphology, and then dedicating separate emulators to each profile shape, or using data augmentation to balance the PCA more across different profile morphologies.  
    
    This stellar profile emulator implementation is publicly available as part of the \texttt{POSYDON} interpolation module along with the binary evolution grids and the \texttt{IFInterpolator} submodule that performs scalar interpolation.\footnote{https://github.com/POSYDON-code/POSYDON/} Although our interpolation methods are currently trained on the grids presented in \texttt{POSYDON} v1 \citep{posydonv1}, we are working to apply them to the forthcoming v2 grids as well. The work presented here provides a framework paving the way for interpolators of array-like quantities; future work will be necessary to optimize the emulation process for specific astrophysical applications. 

\section*{Acknowledgements}
    We thank Tassos Fragos for his input in the conception of this project and his feedback on the design. We thank Mike Zevin for valuable discussion regarding the validation of our models. 

    This work is supported primarily by two sources: the Gordon and Betty Moore Foundation (PI Kalogera, grant awards GBMF8477 and GBMF12341) and the Swiss National Science Foundation (PI Tassos Fragos, project numbers PP00P2\_211006 and CRSII5\_213497).  
    E.T., P.M.S., K.A.R., and M.S. were supported by the project numbers GBMF8477 and GBMF12341.
    V.K. and A.K. gratefully acknowledge the support of the NSF-Simons AI-Institute for the Sky (SkAI) via grants NSF AST-2421845 and Simons Foundation MPS-AI-00010513. 
    Z.D. is grateful for support from the CIERA Board of Visitors Research Professorship.
    J.J.A.~acknowledges support for Program number (JWST-AR-04369.001-A) provided through a grant from the STScI under NASA contract NAS5-03127. 
    S.S.B., M.U.K., and Z.X. were supported by the project number PP00P2\_211006. 
    S.S.B. and M.M.B. were supported by the project number CRSII5\_213497. 
    M.M.B. is also supported by the Boninchi Foundation and the Swiss Government Excellence Scholarship. 
    K.K. is supported by a fellowship program at the Institute of Space Sciences (ICE-CSIC) funded by the program Unidad de Excelencia Mar\'ia de Maeztu CEX2020-001058-M. 
    Z.X. acknowledges support from the Chinese Scholarship Council (CSC). 
    E.Z. acknowledges support from the Hellenic Foundation for Research and Innovation (H.F.R.I.) under the “3rd Call for H.F.R.I. Research Projects to support Post-Doctoral Researchers” (Project No: 7933). 
    The computations were performed at Northwestern University on the Trident computer cluster (funded by the GBMF8477 award) and at the University of Geneva on the Yggdrasil computer cluster. This research was partly supported by the computational resources and staff contributions provided for the Quest high-performance computing facility at Northwestern University, jointly supported by the Office of the Provost, the Office for Research and Northwestern University Information Technology. 

\section*{Software}
    \texttt{numpy} \citep{numpy},
    \texttt{pandas} \citep{pandas},
    \texttt{matplotlib} \citep{matplotlib},
    \texttt{scikit-learn} \citep{scikit-learn},
    \texttt{tensorflow} \citep{tensorflow},
    \texttt{keras} \citep{chollet2015keras},
    \texttt{MESA} \citep{mesacode},
    \texttt{POSYDON} \citep{posydoncode}

\bibliographystyle{elsarticle-harv} 
\bibliography{biblio_revision}

\end{document}